\documentclass[10pt,a4paper]{article}
\usepackage{jheppub}

\usepackage{epsfig}
\usepackage{latexsym}
\usepackage{amsfonts}
\usepackage{amsmath}
\usepackage{amsthm}
\usepackage{amssymb}
\usepackage{amsbsy}
\usepackage{multirow}
\usepackage{slashed}
\usepackage{amsmath,latexsym,amssymb}
\usepackage{graphicx}
\usepackage{psfrag}
\usepackage[latin1]{inputenc}
\usepackage{nicefrac}
\usepackage[stable]{footmisc}
\usepackage{braket}
\usepackage{rotating}
\usepackage{afterpage}
\usepackage{bbm}

\def\eqn#1{eq.~(\ref{#1})}

\def\be{\begin{equation}}
\def\ee{\end{equation}}
\def\bea{\begin{eqnarray}}
\def\eea{\end{eqnarray}}
\def\ba{\begin{eqnarray}}
\def\ea{\end{eqnarray}}

\def\be{\begin{equation}}
\def\ee{\end{equation}}
\def\bea{\begin{eqnarray}}
\def\eea{\end{eqnarray}}
\def\ba{\begin{array}}
\def\ea{\end{array}}
\def\bd{\begin{displaymath}}
\def\ed{\end{displaymath}}

\def\a{\alpha}
\def\b{\beta}

\def\>{\rangle} 
\def\<{\langle} 
\def\Dsl{D \hskip-.6em \raise1pt\hbox{$ / $ } }

\def\da{{\dot\alpha}}
\def\db{{\dot\beta}}

\newsavebox{\uuunit}
\sbox{\uuunit}
    {\setlength{\unitlength}{0.825em}
     \begin{picture}(0.6,0.7)
        \thinlines
        \put(0,0){\line(1,0){0.5}}
        \put(0.15,0){\line(0,1){0.7}}
        \put(0.35,0){\line(0,1){0.8}}
       \multiput(0.3,0.8)(-0.04,-0.02){12}{\rule{0.5pt}{0.5pt}}
     \end {picture}}

\def\be{\begin{equation}}
\def\ee{\end{equation}}
\def\bea{\begin{eqnarray}}
\def\eea{\end{eqnarray}}

\newcommand{\beq}{\begin{eqnarray}}
\newcommand{\eeq}{\end{eqnarray}}

\def\a{\alpha}
\def\b{\beta}

\def\E {$E_{7(7)}$}

\def\sF{{{ F}\!\!\!\!\hskip.8pt\hbox{\raise1pt\hbox{/}}\,}}
\def\som{{{ \omega}\!\!\!\!\hskip.8pt\hbox{\raise1pt\hbox{/}}\,}}
\def\sJ{{{\rm J}\!\!\!\!\hskip.8pt\hbox{\raise1pt\hbox{/}}\,}}

\def\da{{\dot \alpha}}
\def\db{{\dot \beta}}

\def\a{\alpha}
\def\b{\beta}
\def\ba{\bar \alpha}

\title{{ {  Obstruction to  \E\, Deformation  in N=8 Supergravity} }}
\author[a]{Murat  Gunaydin,}    \author[b]{Renata  Kallosh,} 
\affiliation[b]{Institute for Gravitation and the Cosmos and Department of Physics, Pennsylvania State University, \\ University Park, PA 16802, USA
}
\affiliation[a]{Stanford Institute for Theoretical Physics and Department of Physics, Stanford University,\\ Stanford, CA 94305-4060, USA} 
\emailAdd{murat@phys.psu.edu}  \emailAdd{kallosh@stanford.edu}

\abstract{Candidate counterterms break  Noether-Gaillard-Zumino
 \E\,  current conservation in N=8 supergravity in four dimensions. Bossard and Nicolai proposed a scheme for deforming  the subsector involving vector fields in a Lorentz covariant manner, so as to restore duality. They    argued that there must exist an extension of this deformation to the full theory that  preserves  supersymmetry. We show that it is not possible to deform the maximal supergravity to restore \E\ duality, while maintaining both general covariance and $N=8$ supersymmetry, as was proposed. Deformation of N=8 supergravity  requires   higher spins and multiple gravitons, which presents a concrete obstacle to this proposal. 
 \
 
 \
 
 }

\keywords{Supergravity Theories; Higher Spin Fields } 
\begin{document}
 \maketitle
\section{Introduction}

Recently there was a lot of interest in quantum properties of N = 8 supergravity, stimulated by the surprising discovery of its 3-loop finiteness \cite{Bern:2007hh,Bern:2008pv}. There were several attempts to explain this result, see e.g. \cite{Kallosh:2008mq,Brodel:2009hu,Elvang:2010kc,Bossard:2010bd,Beisert:2010jx}. In particular, it was noticed that all known counterterm candidates in N = 8 supergravity would break the Noether-Gaillard-Zumino \E\,  deformed duality current conservation \cite {Kallosh:2011dp,Kallosh:2011qt}. 

 The current non-conservation argument of \cite {Kallosh:2011dp,Kallosh:2011qt} due to a single counterterm was confirmed in 
 \cite{Bossard:2011ij}. However, Bossard and Nicolai (BN) made a proposal how to fix the problem by deforming  the classical  twisted self-duality constraint in presence of higher derivative terms in the action. In classical theory  with \E\, symmetry  there are 28 independent Maxwell field strengths due to a supersymmetric twisted self-duality constraint \cite{Cremmer:1979up,deWit:1982ig}.
To identify a deformed constraint according to BN, one has to find a manifestly duality invariant higher derivative supersymmetric invariant which depends on a double amount of the Maxwell field strengths and their duals, 56 total in N=8 supergravity. Examples of such deformation of the
classical  twisted self-duality constraint were given in  \cite{Bossard:2011ij} for some non-supersymmetric models.

Using the original  BN proposal it was not possible to recover, via the deformation of the self-duality constraint,  even a simple case of a Born-Infeld deformation of the Maxwell theory. The proposal in  \cite{Bossard:2011ij} was further developed  in \cite{Carrasco:2011jv,Chemissany:2011yv,Broedel:2012gf} where  covariant procedures for perturbative non-linear deformations of duality-invariant theories were established. The 
starting point requires the existence of some `Sources of Deformation' (SoD).  Some cases, like Born-Infeld,  requires even an infinite amount of such terms. Various examples of sources of deformation were given in \cite{Carrasco:2011jv,Chemissany:2011yv,Broedel:2012gf} which resulted in building novel models with duality symmetry. Such new models are now available for the case of Born-Infeld models with higher derivatives \cite{Chemissany:2011yv} and models with $U(1)$ duality and global N=2 supersymmetry \cite{Broedel:2012gf}.

The proposal for the vector part of the deformed twisted self-duality constraint in notation of \cite{Bossard:2011ij} is given by 
\be
F^m + J^m{}_n \tilde F^n= G^{mn} {\delta{\cal I} \over \delta F^n} + \Omega^{mn} {\delta{\cal I} \over  \delta \tilde F^n}\,  ,  \qquad m=1,...,56
\label{BN}\ee
Here for N=8 supergravity $G_{mn}$ is a scalar dependent symmetric metric $G_{mn} \in  E_{7(7)} \subset Sp(56, \mathbb{R})$. $J^m{}_n$ is a `complex structure' ,  $\Omega^{mn}$ is a symplectic form and $\tilde F_{\mu\nu} = {1\over 2\sqrt{-g} }\epsilon_{\mu\nu}{}^{\rho\sigma} F_{\rho\sigma}$ is dual to $F$. A duality doublet $F^m$ consists of two sets of 28 Maxwell field strength's
\be
F^m \equiv (F^{\textsl{a}}, F^{\bar {\textsl{a}} })\, , \qquad \textsl{a}=1,...,28 \qquad  \bar {\textsl{a}} =1,...,28
\label{doublet}\ee
In case of N=8 this equation is manifestly \E\, invariant if the SoD is a duality invariant functional depending on a duality doublet  (\ref{doublet}) where the two sets of 28 vectors are  independent. The vector part of SoD is
\be
{\cal I}(F^m) = {\cal I} [ F^{\textsl{a}}, F^{\bar {\textsl{a}} }]
\label{SoD}\ee
The classical twisted self-duality constraint in absence of deformation is 
\be
F^m + J^m{}_n \tilde F^n=0
\label{cl}\ee
which is a relation expressing one of the 28 via the other, so that the theory has only one set of 28 vectors in agreement with unitarity. To deform it according to (\ref{BN}) the SoD action ${\cal I}[ F^{\textsl{a}}, F^{\bar {\textsl{a}} }] $ has to be differentiated over the set of 56 independent vectors. It was assumed in \cite{Bossard:2011ij} that the supersymmetric version of this proposal for SoD is available and has all other symmetries of the theory, local supersymmetry, general covariance etc.

The existence of the supersymmetric SoD in N=8 supergravity was investigated in  \cite{Kallosh:2012yy} where an attempt was made to construct it. Such a SoD when constrained by the classical supersymmetric twisted self-duality constraint (\ref{cl}) has to reproduce the candidate counterterm of N=8 supergravity proposed in
\cite{Kallosh:1980fi,Howe:1980th,Howe:1981xy}. The relevant procedure corresponds to a promotion of the candidate counterterms depending on 28 vectors  to a role of SoD depending on two sets of  independent 28 vectors, while preserving the general covariance and local supersymmetry of the counterterm.
It was concluded in  \cite{Kallosh:2012yy}  that the existence of a supersymmetric version of such an SoD with 56 independent vectors contradicts the N=8 superspace construction \cite{Brink:1979nt}, \cite{Howe:1981gz} and the relevant solutions of the superspace Bianchi identities. It was suggested in \cite{Kallosh:2012yy} that the existing N=8 superspace has to be  deformed to admit the SoD.

The purpose of this note is to make the next level of effort in this direction and to study all possibilities to construct the N=8 supersymmetric SoD. Our strategy will be the following. We first assume that such an SoD  is available with all required properties including i) a double set of 28 independent vectors, ii) non-linear deformed local supersymmetry and iii) general covariance. To evaluate this assumption against the known facts we will study the relevant SoD at the linearized level first, using the fact that  the linearized N=8 supergravity is based on the representations of $SU(2,2|8)$ superconformal algebra. 

We will list all possibilities to double the number of vectors such that upon the use of the classical twisted self-duality constraint (\ref{cl}) the linearized approximation preserves the $SU(2,2|8)$ symmetry. If one does not enlarge the R-symmetry group we will have two options. Either we take multiple copies of the CPT self-conjugate doubleton supermultiplets which do not admit nonlinear extensions or we have to use supermultiplets with spins higher than $2$ and multiple gravitons.  Enlarging  the $R$-symmetry of the superconformal algebra in general leads to an increase in the number of $Q, S$ supersymmetries. This, in turn, implies that supermultiplets  that contain the doublet of vector fields in the ${\bf 28}$ of $SU(8)$ necessarily involve multiple gravitons as well as  fields of spin greater than 2.

We should stress that at the non-linear level the legitimate N=8 supergravity SoD must preserve general covariance.  
The non-linear general covariant actions for spin higher than 2, expandable around flat backgrounds,  are not known to exist. This fact was established in situations studied in the past, see for example \cite {Aragone:1979hx}  and the recent review of the progress in high-spin theories in \cite{Vasiliev:2011zza}. 
 Here  we will argue that the non-linear completion of the  fully supersymmetric invariant SoD actions  require higher spins and multiple gravitons which have no known nonlinear completions that are generally covariant.  This provides an obstruction  to the BN proposal  \cite{Bossard:2011ij} for the deformation of maximal supergravity consistent with  $N=8$  supersymmetry and deformation of  \E\, duality.

\section{Linearized N=8 supergravity and the representations of $SU(2,2|8)$ superconformal algebra}

The massless unitary supermultiplets of extended Poincar\'e superalgebras were classified fully in the early days of space-time supersymmetry\footnote{ For classification of space-time superalgebras see \cite{Nahm:1977tg} and for a complete list of Poincare supermultiplets see \cite{Strathdee:1986jr}.}. The minimum spin range for the massless unitary supermultiplets of N-extended Poincar\'e superalgebras is $\frac{N}{4}$ for even $N$. The shortest supermultiplets for even $N$ are the CPT self-conjugate ones and they contain $N(N-1)/2$ vector fields. Therefore to have a minimum of 56 vector fields  requires that $N> 11$ which means that the corresponding super multiplets have  spin range greater that two.

Since the superfields used in writing down linearized counterms in $N=8$ supergravity correspond to conformal supermultiplets we shall restrict our analysis to conformal superalgebras $SU(2,2|N)$ in four dimensions. 

The oscillator construction of the unitary supermultiplets of extended superconformal algebras in four dimensions  were first given in \cite{Gunaydin:1984fk,Gunaydin:1984vz}  and further developed in \cite{Gunaydin:1998sw,Gunaydin:1998jc}.\footnote{ For  classification of unitary representations of $SU(2,2|1)$ see \cite{Flato:1983te} and for $SU(2,2|N)$ for $N > 1$ see \cite{Dobrev:1985qv}.}

For maximal supergravity the relevant unitary representation is  
the CPT-self-conjugate doubleton supermultiplet of $SU(2,2|8)$. It is obtained by choosing 
the Fock vacuum , $|0\rangle$ , as the lowest weight vector  in the 
$SU(2|4)\times SU(2|4) \times U(1)$ basis in the oscillator  construction ; it was given in  \cite{Gunaydin:1984vz} and we reproduce it in Table \ref{Table_GM}. This supermultiplet  is the $N=8$ counterpart of the Yang-Mills supermultiplet of $SU(2,2|4)$ \cite{Gunaydin:1984fk}.

\begin{table}[ht]
\begin{center}
\begin{tabular}{|c|c|c|c|c|}
\hline
   ${ SL(2,\mathbb{C} ) }$ & ${ E_0} $ & $ {  SU(8)}$ & ${ U(1)}$ &{ Fields}
\\ \hline
 $(0,0)$ & 1     & ${\bf 70}$   & 0 &$\phi^{[ijkl]} $
\\ \hline
 $({1 \over 2},0)$ &  ${3 \over 2} $& ${ \bf 56}$ & 1 &$\lambda^{[ijk]}_{+} \Leftrightarrow \lambda_{\alpha}^{[ijk]}$
\\ \hline
 $(0,{1 \over 2})$   & ${3 \over 2}$   &$\overline{ \bf 56}$ & -1 &$\lambda_{-[ijk]} \Leftrightarrow \lambda_{\dot{\alpha}[ijk]}$
\\ \hline
 (1,0)  & 2 & ${\bf 28}$& 2 &$F_{\mu\nu}^{+[ij]} \Leftrightarrow F_{(\alpha\beta)}^{[ij]}$
\\ \hline
 (0,1)  & 2 & $\overline{\bf 28}$& -2 &$F_{\mu\nu [ij]}^{-} \Leftrightarrow F_{(\dot{\alpha}\dot{\beta})[ij]}$
\\ \hline
 $({3 \over 2},0)$  & ${5 \over 2}$ & ${\bf 8}$& 3 &$\partial_{[ \mu}\psi_{\nu]}^{+ i} \Leftrightarrow \psi_{(\alpha\beta\gamma)}^{i}$
\\ \hline
 $(0,{3 \over 2})$  & ${5 \over 2}$ & $\bar{\bf 8}$& -3 &$\partial_{[ \mu}\psi_{\nu] i}^{-} \Leftrightarrow \psi_{(\dot{\alpha}\dot{\beta}\dot{\gamma}) i}$
\\ \hline
 $(2,0)$  &  3 & $1 $& 4 &$ R_{(\alpha\beta\gamma\delta)}$
\\ \hline
 $(0,2)$  & 3 & $1 $& -4 &$ R_{(\dot{\alpha}\dot{\beta}\dot{\gamma}\dot{\delta} )}$
\\ \hline
\end{tabular}
\medskip
\caption{\small \label{Table_GM} 
The CPT-self-conjugate doubleton supermultiplet of $SU(2,2|8)$ that describe the fields of $N=8$ supergravity in four dimensions. 
  The oscillator formalism gives directly the gauge-invariant field 
strengths associated with the fields in the representation. $i,j,k,..=1,2,..,8 $ are the SU(8) $R$-symmetry  indices. 
The first column indicates the $SL(2,\mathbb{C})$ transformation properties of the field strengths. Columns 3 and 4 list their $SU(8)$ and $U(1)$ transformation properties.
The indices $\alpha, \beta,..$ and $\da, \db, ..$ denote the chiral and anti-chiral spinorial indices of $SL(2,\mathbb{C})$. Round (square) brackets indicate symmetrization (antisymmetrization) of  the enclosed indices.}
\end{center}
\end{table}
Note that the twisted linear self-duality constraint here means that $F_{\mu\nu}^{+[ij]} \Leftrightarrow F_{(\alpha\beta)}^{[ij]}$ is in ${\bf 28}$ whereas $F_{\mu\nu [ij]}^{-} \Leftrightarrow F_{(\dot{\alpha}\dot{\beta})[ij]}$ is in $\overline {\bf 28}$. Relaxing this linear constraint in SoD means that one has to add field strengths $F_{\mu\nu [ij] }^{+} \Leftrightarrow F_{(\alpha\beta) [ij]}$ and $F_{\mu\nu }^{- [ij]} \Leftrightarrow F_{(\dot{\alpha}\dot{\beta})}^{[ij]}$ that transform in the $\overline{\bf 28}$ and ${\bf 28}$ of $SU(8)$, respectively. These additional field strengths  will  eventually be solved as non-linear functions of the original ones, according to \eqn{BN}. But first one has to find the SoD as in \eqn{SoD} depending on a double set of independent vectors which is also supersymmetric and generally covariant, and differentiate it with respect to $F$ to find the deformation providing the rhs of  \eqn{BN}.

The fields of linearized $N=8$ supergravity in four dimensions, which satisfy massless free field equations and massless representations of Poincare group, lift uniquely to those of the conformal group \cite{Mack:1969dg}.  However the nonlinear interactions  of maximal supergravity
 break the conformal supersymmetry algebra $SU(2,2|8)$ down to $N=8$ Poincar\'e superalgebra. As pointed out in \cite{Chiodaroli:2011pp} the superfields used in the construction of counter terms for $N=8$ supergravity \cite{Kallosh:1980fi,Howe:1980th,Howe:1981xy} correspond to the above unitary supermultiplet constructed back in 1984  \cite{Gunaydin:1984vz}. 
This is a consequence of the fact that  the CPT-self-conjugate doubleton supermultiplet of $SU(2,2|8)$ can be assembled into a scalar 
superfield $W_{ijkl}$   depending on  $N=8$ superspace coordinates,
$ \big( x^{\a \dot \b}, \theta^{\a i} , \bar \theta^{\dot \b}_i   \big) $ 
with $\a,\dot \b= 1,2 $ and $i=1,2,\dots 8$. The superfield $W_{ijkl}$ obeys the self-duality condition and the differential constraints 
\be \bar{W}^{ijkl} = {1 \over 4 !} \epsilon^{ijklmnpq} W_{mnpq}\, , \quad \bar{D}_{\dot \a}^i W_{jklm} + {4 \over 5} \delta^i_{[j} \bar{D}^f_{\dot  \a} W_{klm]f} = 0 \ , \quad D_{\alpha i}  W_{jklm}=  D_{\alpha [i}  W_{jklm]}   \label{diffcon4D} \ee

Motivated by the unexpected cancellation of divergences in $N=8$ supergravity up to four loops the superalgebra $SU(2,2|8)$ and the  doubleton supermultiplet of  \cite{Gunaydin:1984vz} have  also been used  more recently  in the construction and analysis of  higher-loop counterterms in maximal
supergravity 
\cite{Beisert:2010jx},\cite{Drummond:2010fp},\cite{Freedman:2011uc}.

\subsection{Examples}

We will start our analysis with interesting examples and later extract the generic features of all possibilities to construct the linearized SoD for N=8 supergravity.

Let us first study the question whether  one can use  the doubleton supermultiplets of $SU(2,2|8)$ as sources of deformation.
First possibility is to consider another copy of CPT self-conjugate doubleton supermultiplet of $SU(2,2|8)$ as SoD. This corresponds to doubling of the fields of $N=8$ supergravity.  It is a long-established fact that ``matter" coupled fully interacting nonlinear supergravity theories exist only for $N\leq 4$. Hence this possibility is  ruled out.
There are other massless supermultiplets of $SU(2,2|8)$ that contain  28 or more vectors. They are obtained very simply via the oscillator method by choosing as the lowest weight vector other states than the vacuum. They are not CPT self-conjugate. Therefore one has to pair them with their CPT conjugate supermultiplets. The full list with 28 or more vectors is given in Tables \ref{Table_GK2}-\ref{Table_GK3}.

\begin{table}[ht]
\begin{center}
\begin{tabular}{|c|c|c|}
\hline
   ${ SL(2,\mathbb{C} ) }$ &  $ {  SU(8)}$  &{ Fields}
\\ \hline
 $(1/2,0)$ & ${\bf 70}$   & $\lambda^{[ijkl]}_+ $
\\ \hline
$({1},0)$ & ${ \bf 56}$ & $~~F^{+[ijk]}_{\mu\nu} \Leftrightarrow F_{(\alpha\beta)}^{[ijk]}~~$
\\ \hline
 $({3/2},0)$ & ${ \bf 28}$ & $\psi^{[ij]}_{\alpha\beta\gamma} $ \\ \hline
  $({2},0)$ & ${ \bf 8}$ & $R^i_{(\alpha\beta\gamma\delta)}$\\ \hline
  $({5/2},0)$ & ${ \bf 1}$ & $R_{(\alpha\beta\gamma\delta\epsilon)}$\\ \hline
 $(0,0)$   &$\overline{ \bf 56}$ & $\phi_{[ijk]} $
\\ \hline
 (0,1/2)  &  $\overline{{\bf 28}}$& $\lambda_{\dot{\alpha} [ij]} $
\\ \hline
 (0,1)  & $\overline{\bf 8}$& $F_{\mu\nu [i]}^{-} \Leftrightarrow F_{(\dot{\alpha}\dot{\beta})[i]}$
\\ \hline
  $(0,{3 \over 2})$  &  ${\bf 1}$& $\partial_{[ \mu}\psi_{\nu] }^{-} \Leftrightarrow \psi_{(\dot{\alpha}\dot{\beta}\dot{\gamma}) }$
\\ \hline
\end{tabular}
\hskip 1.5 cm
\begin{tabular}{|c|c|c|}
\hline
   ${ SL(2,\mathbb{C} ) }$ &  $ {  SU(8)}$  &{ Fields}
\\ \hline
 $(1,0)$ & ${\bf 70}$   & $F^{+[ijkl]}_{\mu\nu} \Leftrightarrow F_{(\alpha\beta)}^{[ijkl]}$
\\ \hline
 $({3/2},0)$ & ${ \bf 56}$ & $\psi^{[ijk]}_{(\alpha\beta\gamma)} $
\\ \hline
 $({2},0)$ & ${ \bf 28}$ & $R^{[ij]}_{(\alpha\beta\gamma\delta)}$ \\ \hline
  $({5/2},0)$ & ${ \bf 8}$ & $\psi^i_{(\alpha\beta\gamma\delta\epsilon)}$\\ \hline
  $({3},0)$ & ${ \bf 1}$ & $R_{(\alpha\beta\gamma\delta\epsilon\kappa)}$\\ \hline
 $(1/2,0)$   &$\overline{ \bf 56}$ & $\lambda_{\alpha[ijk]} $
\\ \hline
 (0,0)  &  $\overline{{\bf 28}}$&$ \phi_{[ij]} $
\\ \hline
 (0,1/2)  & $\overline{\bf 8}$& $\lambda_{\dot{\alpha}i} $
\\ \hline
  $(0,1)$  &  ${\bf 1}$& $F_{\mu\nu}^- \Leftrightarrow F_{(\dot{\alpha}\dot{\beta)}}$
\\ \hline
\end{tabular}
\medskip
\caption{\small \label{Table_GK2} 
The  irreducible chiral doubleton supermultiplets of $SU(2,2|8)$: the one on the left panel has the highest spin $5/2$, the one on the right panel has the highest spin $3$.  }
\end{center}
\end{table}

\begin{table}[ht]
\begin{center}
\begin{tabular}{|c|c|c|}
\hline
   ${ SL(2,\mathbb{C} ) }$ &  $ {  SU(8)}$  &{ Fields}
\\ \hline
 $(3/2,0)$ & ${\bf 70}$   & $\psi^{[ijkl]}_{(\alpha\beta\gamma)} $
\\ \hline
 $({2},0)$ & ${ \bf 56}$ & $R^{[ijk]}_{(\alpha\beta\gamma\delta)}$
\\ \hline
 $({5/2},0)$ & ${ \bf 28}$ & $\psi^{[ij]}_{(\alpha\beta\gamma\delta\epsilon)}$ \\ \hline
  $({3},0)$ & ${ \bf 8}$ & $R^i_{(\alpha\beta\gamma\delta\epsilon\kappa)}$\\ \hline
  $({7/2},0)$ & ${ \bf 1}$ & $\psi_{(\alpha\beta\gamma\delta\epsilon\kappa\lambda)}$\\ \hline
 $(1,0)$   &$\overline{ \bf 56}$ & $F^+_{\mu\nu[ijk]}\Leftrightarrow F_{(\alpha\beta)[ijk]}$
\\ \hline
 (1/2,0)  &  $\overline{{\bf 28}}$&$ \lambda_{\alpha[ij]} $
\\ \hline
 (0,0)  & $\overline{\bf 8}$& $\phi_i $
\\ \hline
  $(0,1/2)$  &  ${\bf 1}$& $\lambda_{\dot{\alpha}}$
\\ \hline
\end{tabular}
\hskip 1.5 cm \begin{tabular}{|c|c|c|}
\hline
   ${ SL(2,\mathbb{C} ) }$ &  $ {  SU(8)}$  &{ Fields}
\\ \hline
 $(2,0)$ & ${\bf 70}$   & $R_{(\alpha\beta\gamma\delta)} $
\\ \hline
 $(5/2,0)$ & ${ \bf 56}$ & $\psi^{[ijk]}_{(\alpha\beta\gamma\delta\epsilon)}$
\\ \hline
 $({3},0)$ & ${ \bf 28}$ & $R^{[ij]}_{(\alpha\beta\gamma\delta\epsilon\lambda)}$ \\ \hline
  $({7/2},0)$ & ${ \bf 8}$ & $\psi^i_{(\alpha\beta\gamma\delta\epsilon\kappa\lambda)}$\\ \hline
  $({4},0)$ & ${ \bf 1}$ & $R_{(\alpha\beta\gamma\delta\epsilon\kappa\lambda\sigma)}$\\ \hline
 $(3/2,0)$   &$\overline{ \bf 56}$ & $\psi_{(\dot{\alpha}\dot{\beta}\dot{\gamma}) [ijk]}$
\\ \hline
 (1,0)  &  $\overline{{\bf 28}}$&$ F_{\mu\nu[ij]}^- \Leftrightarrow F_{(\dot{\alpha} \dot{\beta})[ij]} $
\\ \hline
 (1/2,0)  & $\overline{\bf 8}$& $\lambda_{\alpha i} $
\\ \hline
  $(0,0)$  &  ${\bf 1}$& $\phi$
\\ \hline
\end{tabular}
\medskip
\caption{\small \label{Table_GK3} 
The  irreducible chiral doubleton supermultiplets of $SU(2,2|8)$: the one on the left panel has the highest spin $7/2$, the one on the right panel has the highest spin $4$. }
\end{center}
\end{table}

It is clear from Tables \ref{Table_GK2}-\ref{Table_GK3}  that the other supermultiplets with  28 or more vectors  all break the spin 2 barrier and
furthermore with one exception the vector  fields in them do not transform as {\bf 28} ({\bf $\overline{28}$}) of $SU(8)$. The exception is the supermultiplet given in the right panel in Table \ref{Table_GK3} which has 70 gravitons and spin range of $4$.

Let us now consider unitary supermultiplets of $SU(2,2|8+n)$ with $n\geq 1$ and see if they could be used as SoD's.
The minimal CPT self-conjugate unitary supermultiplet that contains 56 vector fields is the doubleton supermultiplet of $SU(2,2|10)$ which we give in  Table \ref{Table_GK}. This supermultiplet has spin range of $5/2$ and contains 120 vector fields transforming in the antisymmetric tensor of rank three of $SU(10)$. Under the $SU(8)\times SU(2)$ subgroup 120 of $SU(10)$ decomposes as 
\[ 120 =(28,2)+(56,1) + (8,1) \]
Thus we have here the double set of 28 vectors and may study  the non-linear completion of this model. We have 10 gravitons, 45   gravitinos and 252 scalars and a  spin $s=5/2$.  We may note here that in this particular example the doubling of vectors leads to proliferation of gravitons and gravitinos and fields whose spins cross the barrier of $s=2$. 

In the last column of Table \ref{Table_GK} we give the decomposition of the representations of $SU(10)$ with respect to its subgroup $SU(8)\times SU(2)$. If one truncates this unitary supermultiplet of $SU(2,2|10)$ by  throwing out all the $SU(2)$ singlet states one gets two copies of the  CPT self-conjugate supermultiplet of $SU(2,2|8)$. 

The CPT self-conjugate unitary supermultiplet of $SU(2,2|8+2n)$ for $n>0$ contains $\frac{(2n)!}{n!}$ pairs of vector field strength multiplets transforming in $(28+\overline{28})$ of $SU(8)$ subgroup. As such they are candidates for sourcing the deformations of $N=8$ supergravity. However these supermultiplets have spin range of $2+\frac{n}{2}$ and hence break the spin 2 barrier. 

\begin{table}[ht]
\begin{center}
\begin{tabular}{|c|c|c|c|c|c|}
\hline
   ${ SL(2,\mathbb{C} ) }$ & ${ E_0} $ & $ {  SU(10)}$ & ${ U(1)}$ &{ Fields}
& $SU(8)\times SU(2)$ \\ \hline
 $(0,0)$ & 1     & ${\bf 252}$   & 0 &$\phi^{[ijklm]} $ &$(56,1)+(\overline{56},1)+ (70,2)$
\\ \hline
 $({1 \over 2},0)$ &  ${3 \over 2} $& ${ \bf 210}$ & 1 &$\lambda^{[ijkl]}_{+} \equiv \lambda_{\alpha}^{[ijkl]}$ & $(70,1) +(28,1)+(56,2)$
\\ \hline
 $(0,{1 \over 2})$   & ${3 \over 2}$   &$\overline{ \bf 210}$ & -1 &$\lambda_{-[ijkl]} \equiv \lambda_{\dot{\alpha}[ijkl]}$ &$(70,1) +(\overline{28},1)+(\overline{56},2)$
\\ \hline
 (1,0)  & 2 & ${\bf 120}$& 2 &$F_{\mu\nu}^{+[ijk]} \equiv F_{(\alpha\beta)}^{[ijk]}$ &$(56,1)+(8,1)+(28,2)$
\\ \hline
 (0,1)  & 2 & $\overline{\bf 120}$& -2 &$F_{\mu\nu [ijk]}^{-} \equiv F_{(\dot{\alpha}\dot{\beta})[ijk]}$ &$(\overline{56},1)+(\overline{8},1)+(\overline{28},2)$
\\ \hline
 $({3 \over 2},0)$  & ${5 \over 2}$ & ${\bf 45}$& 3 &$\partial_{[ \mu}\psi_{\nu]}^{+ [ij]} \equiv \psi_{(\alpha\beta\gamma)}^{[ij]}$ & $(28,1)+(1,1)+(8,2)$
\\ \hline
 $(0,{3 \over 2})$  & ${5 \over 2}$ & $\bar{\bf 45}$& -3 &$\partial_{[ \mu}\psi_{\nu] [ij]}^{-} \equiv \psi_{(\dot{\alpha}\dot{\beta}\dot{\gamma}) [ij]}$ &$(\overline{28},1)+(1,1)+(\overline{8},2)$
\\ \hline
 $(2,0)$  &  3 & ${\bf10} $& 4 &$ R^i_{(\alpha\beta\gamma\delta)}$ & $(8,1)+(1,2)$
\\ \hline
 $(0,2)$  & 3 & $\overline{\bf 10} $& -4 &$ R_{(\dot{\alpha}\dot{\beta}\dot{\gamma}\dot{\delta}) i}$ &$(\overline{8},1)+(1,2)$
\\ \hline
$({5 \over 2},0)$ & ${7 \over 2}$& 1 & 5 & $ R_{(\alpha\beta\gamma\delta\epsilon)}$ & $(1,1)$ \\ \hline
$(0, {5 \over 2})$ & ${7 \over 2}$& 1 & 5 & $ R_{(\dot{\alpha}\dot{\beta}\dot{\gamma}\dot{\delta}\dot{\epsilon})}$ &$(1,1)$ \\ \hline
\end{tabular}
\medskip
\caption{\small \label{Table_GK} 
An example of the CPT-self-conjugate doubleton supermultiplet of $SU(2,2|10)$.    $i,j,k,..=1,2,..,10 $ are the $SU(10)$ $R$-symmetry  indices. 
}
\end{center}
\end{table}

\subsection{General case} 

A SoD action of  \cite{Bossard:2011ij} is an auxiliary action, whose variation over the duality doublet is designed to provide the deformation of the twisted self-duality constraint. One can argue that the restriction to the unitary supermultiplets may or may not be necessary. The unitarity of the physical action after the deformed twisted self-duality constraint is imposed is, of course, a requirement. However, we may study  more general representations of the enlarged superconformal agebra and not necessarily constrain our analysis to  unitary representations. In such a general case we can still argue that doubling of N=8 vectors while preserving supersymmetry requires higher spins
and multiple gravitons.

A simplest way to argue that doubling of vectors in N=2 supergravity {\it generically} requires  higher spins, $s>2$,  is to look at the symbolic  block diagonal structure of the $SU(2,2|8)$ superconformal algebra \cite{Ferrara:1977ij}. 
\be
\left[\begin{array}{c|c}\rm {conformal \; algebra} & Q, S \\\hline Q, S & R-\rm{symmetry}\end{array}\right]
\label{algebra}\ee

The generators of $SU(2,2|8)$  consist of the 15 generators $P, K, M$  and $D$ of the ordinary conformal group, together with the 64 $R$-symmetry generators of $U(1)$ and  $SU(8)$, and finally, the 64 spinorial charges $Q$  and $S$.
The commutators of $Q$ supersymmetry with $S$ supersymmetry close into  $R$-symmetry, $U(8)$ , Lorentz group and dilatation generators.  
 Physical vectors are defined so that the state of a positive  helicity is in ${\bf 28}$ of $SU(8)$ and the one of  negative helicity is in $\overline {\bf 28}$ as in  Table 1.

If the $R$-symmetry group is not enlarged, the doubling of vectors is possible in special cases, described in Tables \ref{Table_GK2}-\ref{Table_GK3}. In general situation the doubling of vectors can be achieved by enlarging the $R$-symmetry of the models as we saw above. 
Once the lower right corner of the algebra matrix (\ref{algebra}) corresponding to R-symmetry generators is enlarged in any  way, one has to increase the numbers of $Q$ and $S$ type supersymmetry generators.  When  $Q$ and $S$ carry internal indices that run from 1 to  $8+n$ with $n>1$, one crosses the barrier of spin  2. For  Poincar\'e supersymmetry this follows from the results of  \cite{Nahm:1977tg} and in the superconformal case from those of  \cite{Gunaydin:1984vz}.

\section{A Non-linear Supersymmetric Completion of the Source of Deformation of \E\,  :   No-Go }

The reason for us to use the linearized approximation was to control supersymmetry, which was relatively easy at the linear level. We were able to provide a list of all  possibilities to double the vector states consistent with linearized supersymmetry.  However, at the linear level the theory has nothing to say about the \E\, symmetry and about the scalars forming coset spaces or any other geometries.  Meanwhile, at the non-linear level vectors and scalars of N=8 supergravity transform under \E\,  and moreover, the SoD is expected to preserve general covariance.

To summarize we have shown that  doubling the vectors in N=8 supergravity for the purpose of producing SoD, while preserving supersymmetry,   can be achieved in two different ways.
\begin{itemize}
  \item  At the linear level one can consider a  {\it double or multiple copies of N=8 supergravity multiplets}  as a candidates for the SoD. The superposition principle allows this. However, these models do not have a non-linear completion. An N=8 supergravity multiplet has only a self-coupling.   
     
  \item We have discussed  all other possibilities to double the vectors in N=8 supergravity in agreement with supersymmetry, and we have shown     that {\it in all cases the corresponding linearized models cross the barrier of $s=2$ and have multiple gravitons}. 
\end{itemize}

One might contemplate a  scenario of  N=8 supergravity superspace deformation
\footnote{ We are grateful to H. Nicolai and P. Howe for a discussion of such a scenario.  It does not rely on a Lorentz non-covariant version of the deformation of N=8 proposal \cite{Bossard:2010dq,Bossard:2011ij}  based on  a non-covariant Henneaux-Teitelboim formalism \cite{Henneaux:1988gg,Bunster:2012km}.
 } which would  avoid the contradiction discussed in  \cite{Kallosh:2012yy} of the duality deformation proposal  \cite{Bossard:2011ij} with the standard superspace Bianchi Identities in \cite{Brink:1979nt}, \cite{Howe:1981gz}. 
In this scenario one modifies  the 
superspace torsion constraints and  introduces
superfields for all the component fields including the additional vector fields with the exception  of the  vierbein and gravitino that are still part of the standard   superspace vielbein. In particular, 
 56 vector field strengths would be elevated to superfields, as
would be the linear twisted selfduality constraint.  
Our investigation of the linearized version of the SoD for N=8 supergravity shows that such a  deformation of superspace  is not possible  in principle. We studied all supermultiplets involving 2 sets of 28 vectors and concluded that higher spins and multiple gravitons are inevitable.

With regard to higher spins and multiple gravitons, required to deform N=8 supergravity, one should take into account certain facts\footnote{We are grateful to M. Vasiliev for updating us on the current status of higher-spin field theories.}. The 
 old result  of Aragone and Deser \cite {Aragone:1979hx}  
 is still valid \cite{Vasiliev:2011zza}, namely there are
 no  local interactions of massless
higher-spin fields with gravity which can be consistently
expanded around Minkowski background. Furthermore,  in general higher-spin theory, fields of different spins (including spin two) may carry color indices of Chan-Paton type when spin 1 massless fields belong to some non-Abelian gauge group. Graviton in these models is a 
singlet (colorless) spin two field \cite{Vasiliev:1988sa,Konstein:1989ij}. 

Now we may evaluate the proposal \cite{Bossard:2011ij} to deform N=8 supergravity in view of these properties of the higher-spin models and the results presented above. 
The counterterms of N=8 supergravity which we try to promote to SoD  have a 4-graviton coupling, a 4-gravitino coupling, a 4-vector coupling etc \footnote{See for example all details for $R^4$ in \cite{Freedman:2011uc} in eq. (6.8).}. The contribution of candidate counterterms to the 4-graviton coupling is proportional to the tree level 4-graviton coupling times some polynomial of $s, t, u$,  whose dimension depends on the loop number, for  counterterms of the form  $D^{2(L-3)} R^4$. For the 3-loop case there is a factor $stu$ times the tree amplitude.
To promote the counterterms to the level of SoD, depending on a double set of vectors one may use any of the supermultiplets described above, which  involve $s>2$ and multiple gravitons. To the total  superamplitude of  physical particles in N=8 supergravity we have to add the gauge-invariant 4-couplings of higher-spin particles with $N>8$ supersymmetries. Higher-order vertices of this type (usually called Abelian)
were recently discussed in \cite{Ruehl:2011tk,Vasilev:2011xf}, however, they are not supersymmetric. Whether such couplings can be made supersymmetric remains an open problem. 
    Thus already at the linearized level the inclusion of  4-point amplitudes of higher spins presents itself as a concrete obstacle to the supersymmetric extension of the proposal of \cite{Bossard:2011ij}.

The non-linear local completion of the linearized SoD (ignoring  the 4-coupling issue of higher spins) has a general covariance problem of the Aragone-Deser type. Moreover, it is aggravated by the multiple gravitons. The choice of the singlet for a non-Abelian group on vectors, which may help in some high spin models  is not helpful here. A direct inspection of our tables shows that the gravitons are not singlets in $SU(8)$ (except the case with double copies of N=8 supergravity, which has no non-linear completion anyway)\footnote{ There exist supermultiplets of $SU(2,2|8+n)$ with the graviton transforming as a singlet other than the CPT self-conjugate doubleton of $SU(2,2|8)$. However they do not contain vector fields.}. This is another concrete obstacle to N=8 supergravity deformation. The concept of general covariance has no clear meaning in this situation where gravitons form, for example, 70 of $SU(8)$ as in the rhs of the Table \ref{Table_GK3}  and vectors form 2 sets of 28 of $SU(8)$, as required for the SoD.

Clearly higher spins and multiple  gravitons present an obstacle to the superspace deformation scenario discussed above, which would rescue the supersymmetric N=8 deformation proposal \cite{Bossard:2011ij}. For example, the main cornestone of the superspace geometry, the supervielbein which is a bridge between $SU(8)$ and \E\,, is lost when we have to deal with multiple gravitons required for supersymmetry. 

The proposal of  \cite{Bossard:2011ij} met with  success in cases of novel models with global  N=2 supersymmetry \cite{Broedel:2012gf}. It is therefore interesting to compare and explain why it fails in N=8 supergravity. The reason why things work well for global N=2 supersymmetry and duality is because in global supersymmetry it is easy to double the set of N=2 matter superfields, and make them non-linearly interacting, as demonstrated in \cite{Broedel:2012gf}, where numerous examples of SoD's were presented. In N=8 supergravity there is one CPT conjugate supermultiplet and it is self-interacting. Doubling this unique supermultiplet is possible at the linear level, however, these two copies of N=8 supergravity do not interact. Another way to double the vectors of the N=8 supergravity multiplet is to deform the superconformal R-symmetry of the linearized supersymmetry.  This in turn leads  to a crossing of the $N=8$, $s=2$ barrier and to multiple gravitons, which makes it impossible to implement  the supersymmetric extension  of the non-linear twisted self-duality  proposal of \cite{Bossard:2011ij} within the framework of  N=8 supergravity.

\section{Acknowledgement}
We are grateful to J. J. M. Carrasco, P. Howe,  H. Nicolai,  A. Linde, T. Ortin, R. Roiban and M. Vasiliev  for the stimulating discussions. RK is supported by Stanford Institute for Theoretical Physics (SITP), the NSF Grant No. 0756174 and the John Templeton foundation grant `Quantum Gravity Frontiers'. The work of MG is supported by US National Science Foundation under grants PHY-12-13183 and PHY-08-55356  and he  acknowledges the hospitality of  SITP where part of this work was performed.

\providecommand{\href}[2]{#2}\begingroup\raggedright\endgroup



\begin{thebibliography}{10}

\bibitem{Bern:2007hh}
Z.~Bern, J.~Carrasco, L.~J. Dixon, H.~Johansson, D.~Kosower {\em et al.},  {\em
  {Three-Loop Superfiniteness of N=8 Supergravity}}, Phys.Rev.Lett. {\bf 98}
  (2007) 161303
[\href{http://www.arXiv.org/abs/hep-th/0702112}{{\tt hep-th/0702112}}].

\bibitem{Bern:2008pv}
Z.~Bern, J.~Carrasco, L.~J. Dixon, H.~Johansson and R.~Roiban,  {\em {Manifest
  Ultraviolet Behavior for the Three-Loop Four-Point Amplitude of N=8
  Supergravity}}, Phys.Rev. {\bf D78} (2008) 105019
[\href{http://www.arXiv.org/abs/0808.4112}{{\tt 0808.4112}}].

\bibitem{Kallosh:2008mq}
R.~Kallosh,  {\em {On a possibility of a UV finite N=8 supergravity}},
\href{http://www.arXiv.org/abs/0808.2310}{{\tt 0808.2310}}.

\bibitem{Brodel:2009hu}
J.~Broedel and L.~J. Dixon,  {\em {$R^4$ counterterm and $E_{7(7)}$  symmetry in
  maximal supergravity}}, JHEP {\bf 1005} (2010) 003
[\href{http://www.arXiv.org/abs/0911.5704}{{\tt 0911.5704}}].

\bibitem{Elvang:2010kc}
H.~Elvang and M.~Kiermaier,  {\em {Stringy KLT relations, global symmetries,
  and $E_{7(7)}$ violation}}, JHEP {\bf 1010} (2010) 108
[\href{http://www.arXiv.org/abs/1007.4813}{{\tt 1007.4813}}].

\bibitem{Bossard:2010bd}
G.~Bossard, P.~Howe and K.~Stelle,  {\em {On duality symmetries of supergravity
  invariants}}, JHEP {\bf 1101} (2011) 020
[\href{http://www.arXiv.org/abs/1009.0743}{{\tt 1009.0743}}].

\bibitem{Beisert:2010jx}
N.~Beisert, H.~Elvang, D.~Z. Freedman, M.~Kiermaier, A.~Morales {\em et al.},
  {\em {$E_{7(7)}$  constraints on counterterms in N=8 supergravity}}, Phys.Lett.
  {\bf B694} (2010) 265--271
[\href{http://www.arXiv.org/abs/1009.1643}{{\tt 1009.1643}}].

\bibitem{Kallosh:2011dp}
R.~Kallosh,  {\em {$E_{7(7)}$ Symmetry and Finiteness of N=8 Supergravity}},
  JHEP {\bf 1203} (2012) 083
[\href{http://www.arXiv.org/abs/1103.4115}{{\tt 1103.4115}}].

\bibitem{Kallosh:2011qt}
R.~Kallosh,  {\em {N=8 Counterterms and $E_{7(7)}$ Current Conservation}}, JHEP
  {\bf 1106} (2011) 073
[\href{http://www.arXiv.org/abs/1104.5480}{{\tt 1104.5480}}].

\bibitem{Bossard:2011ij}
G.~Bossard and H.~Nicolai,  {\em {Counterterms vs. Dualities}}, JHEP {\bf 1108}
  (2011) 074
[\href{http://www.arXiv.org/abs/1105.1273}{{\tt 1105.1273}}].

\bibitem{Cremmer:1979up}
E.~Cremmer and B.~Julia,  {\em {The SO(8) Supergravity}}, Nucl.Phys. {\bf B159}
  (1979)
141.

\bibitem{deWit:1982ig}
B.~de~Wit and H.~Nicolai,  {\em {N=8 Supergravity}}, Nucl.Phys. {\bf B208}
  (1982)
323.

\bibitem{Carrasco:2011jv}
J.~J.~M. Carrasco, R.~Kallosh and R.~Roiban,  {\em {Covariant procedures for
  perturbative non-linear deformation of duality-invariant theories}},
  Phys.Rev. {\bf D85} (2012) 025007
[\href{http://www.arXiv.org/abs/1108.4390}{{\tt 1108.4390}}].

\bibitem{Chemissany:2011yv}
W.~Chemissany, R.~Kallosh and T.~Ortin,  {\em {Born-Infeld with Higher
  Derivatives}}, Phys.Rev. {\bf D85} (2012) 046002
[\href{http://www.arXiv.org/abs/1112.0332}{{\tt 1112.0332}}].

\bibitem{Broedel:2012gf}
J.~Broedel, J.~J.~M. Carrasco, S.~Ferrara, R.~Kallosh and R.~Roiban,  {\em {N=2
  Supersymmetry and U(1)-Duality}}, Phys.Rev. {\bf D85} (2012) 125036
[\href{http://www.arXiv.org/abs/1202.0014}{{\tt 1202.0014}}].

\bibitem{Kallosh:2012yy}
R.~Kallosh and T.~Ortin,  {\em {New $E_{7(7)}$  invariants and amplitudes}}, JHEP {\bf
  1209} (2012) 137
[\href{http://www.arXiv.org/abs/1205.4437}{{\tt 1205.4437}}].

\bibitem{Kallosh:1980fi}
R.~Kallosh,  {\em {Counterterms in extended supergravities}}, Phys.Lett. {\bf
  B99} (1981)
122--127.

\bibitem{Howe:1980th}
P.~S. Howe and U.~Lindstrom,  {\em {Higher order invariants in extended
  supergravity}}, Nucl.Phys. {\bf B181} (1981)
487.

\bibitem{Howe:1981xy}
P.~S. Howe, K.~Stelle and P.~Townsend,  {\em {Superactions}}, Nucl.Phys. {\bf
  B191} (1981)
445.

\bibitem{Brink:1979nt}
L.~Brink and P.~S. Howe,  {\em {The N=8 supergravity in superspace}},
  Phys.Lett. {\bf B88} (1979)
268.

\bibitem{Howe:1981gz}
P.~S. Howe,  {\em {Supergravity in superspace}}, Nucl.Phys. {\bf B199} (1982)
309.

\bibitem{Aragone:1979hx}
C.~Aragone and S.~Deser,  {\em {Consistency Problems of Hypergravity}},
  Phys.Lett. {\bf B86} (1979)
161.

\bibitem{Vasiliev:2011zza}
M.~A. Vasiliev,  {\em {V L Ginzburg and higher-spin fields}}, Phys.Usp. {\bf
  54} (2011)
641--648.

\bibitem{Nahm:1977tg}
W.~Nahm,  {\em {Supersymmetries and their Representations}}, Nucl.Phys. {\bf
  B135} (1978)
149.

\bibitem{Strathdee:1986jr}
J.~Strathdee,  {\em {EXTENDED POINCARE SUPERSYMMETRY}}, Int.J.Mod.Phys. {\bf
  A2} (1987)
273.

\bibitem{Gunaydin:1984fk}
M.~Gunaydin and N.~Marcus,  {\em {The Spectrum of the $S^5$  Compactification of
  the Chiral N=2, D=10 Supergravity and the Unitary Supermultiplets of $U(2,
  2/4)$}}, Class.Quant.Grav. {\bf 2} (1985)
L11.

\bibitem{Gunaydin:1984vz}
M.~Gunaydin and N.~Marcus,  {\em {THE UNITARY SUPERMULTIPLET OF N=8 CONFORMAL
  SUPERALGEBRA INVOLVING FIELDS OF SPIN $\leq 2$}}, Class.Quant.Grav.
  {\bf 2} (1985)
L19.

\bibitem{Gunaydin:1998sw}
M.~Gunaydin, D.~Minic and M.~Zagermann,  {\em {4-D doubleton conformal
  theories, CPT and IIB string on $AdS_5 \times S^5 $}, Nucl.Phys. {\bf B534}
  (1998) 96--120
[\href{http://www.arXiv.org/abs/hep-th/9806042}{{\tt hep-th/9806042}}].

\bibitem{Gunaydin:1998jc}
M.~Gunaydin, D.~Minic and M.~Zagermann,  {\em {Novel supermultiplets of
  $SU(2,2|4)$ and the $AdS_5 / CFT_4$  duality}}, Nucl.Phys. {\bf B544} (1999)
  737--758
[\href{http://www.arXiv.org/abs/hep-th/9810226}{{\tt hep-th/9810226}}].

\bibitem{Flato:1983te}
M.~Flato and C.~Fronsdal,  {\em {REPRESENTATIONS OF CONFORMAL SUPERSYMMETRY}},
  Lett.Math.Phys. {\bf 8} (1984)
159.

\bibitem{Dobrev:1985qv}
V.~Dobrev and V.~Petkova,  {\em {All Positive Energy Unitary Irreducible
  Representations of Extended Conformal Supersymmetry}}, Phys.Lett. {\bf B162}
  (1985)
127--132.

\bibitem{Mack:1969dg}
G.~Mack and I.~Todorov,  {\em {Irreducibility of the ladder representations of
  u(2,2) when restricted to the poincare subgroup}}, J.Math.Phys. {\bf 10}
  (1969)
2078--2085.

\bibitem{Chiodaroli:2011pp}
M.~Chiodaroli, M.~Gunaydin and R.~Roiban,  {\em {Superconformal symmetry and
  maximal supergravity in various dimensions}}, JHEP {\bf 1203} (2012) 093
[\href{http://www.arXiv.org/abs/1108.3085}{{\tt 1108.3085}}].

\bibitem{Drummond:2010fp}
J.~Drummond, P.~Heslop and P.~Howe,  {\em {A Note on N=8 counterterms}},
\href{http://www.arXiv.org/abs/1008.4939}{{\tt 1008.4939}}.

\bibitem{Freedman:2011uc}
D.~Z. Freedman and E.~Tonni,  {\em {The $D^{2k} R^4$ Invariants of N=8
  Supergravity}}, JHEP {\bf 1104} (2011) 006
[\href{http://www.arXiv.org/abs/1101.1672}{{\tt 1101.1672}}].

\bibitem{Ferrara:1977ij}
S.~Ferrara, M.~Kaku, P.~Townsend and P.~van Nieuwenhuizen,  {\em {Gauging the
  Graded Conformal Group with Unitary Internal Symmetries}}, Nucl.Phys. {\bf
  B129} (1977)
125.

\bibitem{Bossard:2010dq}
G.~Bossard, C.~Hillmann and H.~Nicolai,  {\em $E_{7(7)}$  symmetry in perturbatively
  quantised N=8 supergravity}}, JHEP {\bf 1012} (2010) 052
[\href{http://www.arXiv.org/abs/1007.5472}{{\tt 1007.5472}}].

\bibitem{Henneaux:1988gg}
M.~Henneaux and C.~Teitelboim,  {\em {DYNAMICS OF CHIRAL (SELFDUAL) P FORMS}},
  Phys.Lett. {\bf B206} (1988)
650.

\bibitem{Bunster:2012km}
C.~Bunster, M.~Henneaux and S.~Hortner,  {\em {Gravitational Electric-Magnetic
  Duality, Gauge Invariance and Twisted Self-Duality}},
\href{http://www.arXiv.org/abs/1207.1840}{{\tt 1207.1840}}.

\bibitem{Vasiliev:1988sa}
M.~A. Vasiliev,  {\em {CONSISTENT EQUATIONS FOR INTERACTING MASSLESS FIELDS OF
  ALL SPINS IN THE FIRST ORDER IN CURVATURES}}, Annals Phys. {\bf 190} (1989)
59--106.

\bibitem{Konstein:1989ij}
S.~Konstein and M.~A. Vasiliev,  {\em {EXTENDED HIGHER SPIN SUPERALGEBRAS AND
  THEIR MASSLESS REPRESENTATIONS}}, Nucl.Phys. {\bf B331} (1990)
475--499.

\bibitem{Ruehl:2011tk}
W.~Ruehl,  {\em {Solving Noether's equations for gauge invariant local
  Lagrangians of N arbitrary higher even spin fields}},
\href{http://www.arXiv.org/abs/1108.0225}{{\tt 1108.0225}}.

\bibitem{Vasilev:2011xf}
M.~Vasiliev,  {\em {Cubic Vertices for Symmetric Higher-Spin Gauge Fields in
  $(A)dS_d$}}, Nucl.Phys. {\bf B862} (2012) 341--408
[\href{http://www.arXiv.org/abs/1108.5921}{{\tt 1108.5921}}].

\end{thebibliography}

\end{document}